# Fluorescent Carbon Nanoparticle:

# Synthesis, Characterization and Bio-imaging Application

S.C. Ray[(a),*], Arindam Saha, Nikhil **R.** Jana[*] and Rupa Sarka**r**

*Centre for Advanced Materials, Indian Association for the Cultivation of Science, Kolkata-700032 (India)*

**Abstract:** Fluorescent carbon nanoparticle (CNP) having 2-6 nm in size with quantum yield of about ~3% were synthesized via nitric acid oxidation of carbon soot and this approach can be used for milligram scale synthesis of these water soluble particles. These CNPs are nano-crystalline with predominantly graphitic structure and shows green fluorescence under UV exposure. While nitric acid oxidation induces nitrogen and oxygen incorporation into soot particle that afforded water solubility and light emitting property; the isolation of small particles from a mixture of different size particles improved the fluorescence quantum yield. These CNP shows encouraging cell imaging application. They enter into cell without any further functionalization and fluorescence property of these particles can be used for fluorescence based cell imaging application.

---

[(a)]**Present Address**: School of Physics, University of the Witwatersrand, Private Bag 3, Wits 2050, South Africa.

*\*Corresponding author(s):* S.C. Ray (raysekhar@rediffmail.com) and Nikhil R. Jana (camnrj@iacs.res.in)

## Introduction

The emergence of fluorescence carbon nanoparticle (CNP) shows high potential in biological labeling, bioimaging and other different optoelectronic device applications.[1-13] These carbon nanoparticles are biocompatible and chemically inert,[2,6,14-18] which has advantages over conventional cadmium based quantum dots.[19] However, these fluorescent carbon nanoparticle are poorly studied compared to other carbon-based materials such as carbon nanotubes and fullerenes. In addition, the understanding of the origin of fluorescence in carbon nanoparticle is far from sufficient.[4,5,7,9] For example, information on the microstructure and surface ligands remains unclear and details of the organic passivation is not sufficient to aid understanding of the surface states beneficial for light emission.

Common routes in making fluorescent carbon nanoparticle includes high energy ion beam radiation based creation of point defect in diamond particle followed by annealing,[1,3] laser ablation of graphite followed by oxidation and functionalization,[4,7] thermal decomposition of organic compound,[10,11,13] electrooxidation of graphite[9] and oxidation of candle soot with nitric acid.[8] A wide range of fluorescent carbon particle of different colors can be prepared by those approaches; e.g. octadecylamine functionalized diamond nanoparticle showed blue fluorescence,[12] nitrogen doped diamond showed red fluorescence[1a] and candle soot derived particle[8] or thermal decomposition method[11,13] or laser ablation method[4] produced particles with multiple colors. However, quantum yield of most of these particle are too low (<1%)[8,9] with few exceptions.[4,13] In addition the synthetic methods are cumbersome and inefficient. For example, in the high energy ion beam radiation based method, it is difficult to introduce a large number of point defects into ultra-fine nano-carbon particles (<10 nm) for bright luminescence.[1,3] Thermal decomposition based methods produce low yield of soluble and fluorescent particle with a significant fraction of insoluble product.[10,11,13] Soot based synthesis produce particle mixture of different colors and isolation of different colored particles by gel electrophoresis is a difficult task.[8] Recent report showed that surface passivation can lead to a significant increase in fluorescence quantum yield (4-15%) however exact mechanism is not yet clear.[4,13] Thus simple, efficient and large scale synthesis of fluorescent carbon nanoparticle and their isolation, purification and functionalization are very challenging.

Among all these synthetic methods, soot based approach is simple and straightforward.[8] However, quantum yield of fluorescent carbon nanoparticle is too low (<0.1%) to any useful application.[8] Herein, we report an improved soot based method of synthesizing fluorescent carbon nanoparticle (CNP) of 2-6 nm in size with quantum yield of ~3%. There are three distinct improvements in our modified method. First, we developed a simple separation method of small size and fluorescent carbon particle from heterogeneous particle mixture. The method is applicable for milligram scale synthesis of these particles. Second, small particles are more fluorescent than larger one and thus isolation of small particle improves the quantum yield from <0.1% to ~3%. Third, we observed that these small carbon particles enter into cell without any further functionalization and fluorescence property of particle can be used for fluorescence based cell imaging application.

## Experimental Procedures

*Synthesis of carbon particle*

25 mg carbon soot (collected from burning candle) was mixed with 15 mL of 5M nitric acid in a 25 mL three necked flask. It was then refluxed at 100°C for 12 hours with magnetic stirring. After that the black solution was cooled and centrifuged at 3000 rpm for 10 minutes to separate out unreacted carbon soot. The light brownish yellow supernatant was collected that shows green fluorescence under UV exposure. The aqueous supernatant was mixed with acetone (water:acetone volume ratio was 1:3) and centrifuged at 14000 rpm for 10 minutes. The black precipitate was collected and dissolved in 5-10 mL water. The colorless and non-fluorescent supernatant was discarded. This step of purification separates excess nitric acid from the carbon nanoparticles. This concentrated aqueous solution having almost neutral pH was taken for further use. The same synthesis technique was also performed for 6 hours reflux and 18 hours reflux. The supernatant obtained from 6 hours reflux was pale yellow and for 18 hour reflux was dark yellow. We weighted the unreacted carbon soot which was removed as precipitate, in order to find out the yield of soluble carbon nanoparticles. The weight was ~22.5 mg for 6 hours reflux time (yield ~10%) and ~20 mg for 12 and 18 hours reflux time (yield ~20%). It shows that yield increases as reflux time increases from 6 to 12 hours but after that no significant increases was observed. This solution has particles having size ranges from 20-350 nm and called as synthesized carbon particle (CP).

*Size separation of carbon particles*

Size separation was performed in a solvent mixture with the combination of high speed centrifuge based separation. As synthesized carbon particles (CP) are soluble in water, ethanol and acetone but insoluble in chloroform. The aqueous and ethanolic particle solution does not precipitate even at 16000 rpm centrifugation. So we chose a solvent mixture of water-ethanol-chloroform (single phase without any phase separation) for the size separation of particle where water-ethanol helps to solubilize the particle but chloroform decreases their solubility. Next, we followed a step-by-step separation using different centrifugation speed from 4000-16000 rpm.

In a typical process, the aqueous solution of carbon nanoparticle was mixed with chloroform and ethanol maintaining water:chloroform:ethanol volume ratio of 1:1:3. Next, 2 mL of this solution was centrifuged with 4000 rpm for 10 minutes. The precipitate was collected and dissolved in 2 mL fresh water. The supernatant was then again centrifuged at 5000 rpm for 10 minutes and the precipitate was collected and redissolved in 2 mL of water. The same procedure was applied at 6000 rpm and 8000 rpm. The supernatant obtained after 8000 rpm was not getting precipitated even at 16000 rpm. This solution was collected and evaporated to dryness to remove ethanol and chloroform and finally dissolved in water. This solution has particle size of 2-6 nm and named as carbon nanoparticle (CNP) that were used for characterization and application. The yield of CNP from CP is ~ 50%, which means 25 mg soot can produce ~2-3 mg of CNP. In this calculation we assumed that carbon dioxide formation is negligible during nitric acid oxidation.

*Cell labeling and cytotoxicity assay*

For cell labeling experiment, Ehrlich ascites carcinoma cells (EAC) were collected from peritoneal cavity of adult female mice after 7 days of inoculation. The suspension of cells was prepared with a concentration ~ $10^7$ cell / mL. Next, 1.0 mL of this suspension was mixed with 10-100 µL of aqueous CNP solution and incubated for 30 minutes. Next, labelled cells were separated from free CNP by centrifuging at 2000 rpm for 3 minutes. The precipitated cells were suspended in phosphate buffer solution. This type of precipitation-resuspended was repeated for 2 more times and finally cells were suspended in phosphate buffer solution. A drop of this suspension was placed in a glass slide for imaging experiment. Fluorescence image was captured using Olympus IX71 fluorescence microscope with DP70 digital camera.

For MTT and Trypan blue assay, HepG2 cells were trypsinized and resuspended in culture medium. The cells were seeded to flat bottom microplate with 0.5 mL full medium and kept for overnight at 37ºC and 5% $CO_2$. The CNP solution of different amounts was loaded to each well and each concentration has 3 duplications. After incubation for 24 hours, 50µL of MTT solution (5mg/mL) was added to each well 4 hours before the end of the incubation. The medium was discarded and produced formazan was dissolved with DMSO. The plates were read with absorbance at 550nm. The optical density is directly correlated with cell quantity and cell viability was calculated by assuming 100% viability in the control set without any CNP. In case of Trypan blue assay, 0.4% of Trypan blue solution was used instead of MTT and after 5 minutes stained cells are counted to determine the cell viability.

*Instrumentation*

Diluted CNP solution was dropped onto copper grids to prepare specimens for transmission electron microscopic (TEM) observation which was performed in a FEI Tecnai G2 F20 microscope with a field-emission gun operating at 200 kV. The microstructures of the CNPs were examined using JEOL JSM-6700F scanning electron microscope and Veeco di CP-II atomic force microscope respectively. The fluorescence (photoluminescence) spectra were measured on a Hitachi F4500 fluorescence spectrophotometer at different excitation energy ranging from 325 nm to 600 nm. A Nicolet 6700 Fourier transform infrared (FTIR) spectrophotometer was used to analyze chemical bonds on the surface of CNP. A Perkin-Elmer PH1-1600 X-ray photoelectron spectrometer (XPS) was used for compositional analysis and chemical bond determination of CNPs. Dynamic light scattering (DLS) study was performed using model BI-200SM instrument (Brookhaven

Instrument Corporation), after filtering the sample solution with Milipore syringe filter (0.2 micron pore size). Micro Raman studies was performed using an ISA Lab Raman system equipped with 514.5 nm LASER with a 100×objective giving a spot size about 1 μm with a spectral resolution better than 2 cm$^{-1}$. The quantum yield was measured by comparing the integrated photoluminescence intensities and absorbance values of the CNP with the reference fluorescein dye (QY = 95 %).

## Results and discussion

Carbon soot is black in color and completely insoluble in water even after ultra-sonnication. This is because they are large in size and hydrophobic in nature. When this soot is refluxed with nitric acid, light brown colored supernatant solution is obtained along with an insoluble black precipitate. Brownish yellow supernatant indicates a part of carbon particle becomes small and water soluble during the refluxing processes. This soluble particle exhibits green fluorescence when irradiates with UV light, while the precipitate part shows no fluorescence. We have studied the fluorescence spectra at different excitation energy ranging from 325 nm - 600 nm and found that the highest fluorescence intensity was achieved at the excitation wavelength of 450 nm, which shows emission maximum at 520 nm. Figure 1a shows the photoluminescence spectra of as synthesized CP with different excitation wavelength. We have tested the quality and yield of CP as a function of reflux time and found that 12 hour is the optimum time. If the reflux time is less the yield is low and for longer reflux time the yield does not increase appreciably. (See experimental section and supporting Figure S1a).

The nature of fluorescence spectra suggested that there are different types of particles with different colors. Earlier work also showed that these particles can be separated by gel electrophoresis technique.[8] We have performed size separation of CP in order to separate CP of different fluorescent property. The size separation has been performed in a solvent mixture with the combination of high speed centrifuge based separation. We have identified mixture of water-ethanol-chloroform as a single phase solvent for the effective size separation of CP, where water-ethanol helps to solubilize the CP but chloroform decreases their solubility. Next we have followed a step-by-step separation using different centrifugation speed from 4000-16000 rpm.

Figure 1b shows the fluorescence spectra of CP solution after successive size separation. It shows that smallest particle (CNP) that does not precipitate even at 16000 rpm, shows the highest fluorescence. All the other size particles show very weak fluorescence and there is very little blue shift in fluorescence with decreasing particle size. Fluorescence study of various phosphate buffer solution of CNP shows that fluorescent intensity does not change appreciably on solution pH from 7-9. (See in supporting Figure S1b). The comparative absorption spectra of as synthesized carbon particle (CP) and smallest particle (CNP) are shown in Figure 1c. No appreciable change is observed except that CNP shows a stronger absorbance in visible wavelength along with weak band at ~ 350 nm.

Transmission electron microscopic (TEM) study shows that as synthesized carbon particle (CP) have broad size distribution from 20-350 nm but CNP have small and narrow particle size distribution from 2-6 nm (Figure 2a and 2b). Both CP and CNP has well graphitization, the interlayer spacing between graphitic sheets is $d_{(002)}$=0.33 nm (obtained from HRTEM: shown as inset in Figure 2b), which is very close to that of the ideal graphite. Similar type of carbon nanoparticle (1.5 – 2.5 nm) is reported earlier following an aqueous route with the help of silica sphere as carrier.[13] Comparison of TEM and fluorescence data shows that CNP of 2-6 nm size has higher fluorescence intensity compared to CP of larger overall size. This type of size dependent fluorescence QY is observed for carbon particle produced via laser ablation technique,[4] thermal decomposition[11,13] and particle obtained from candle soot.[8]

We found that our CP and CNP has very strong tendency of aggregation during TEM grid preparation or SEM slide preparation. The aggregation is so high that we face difficulty in finding significant amount of isolated small carbon particle (CNP) under TEM either from as synthesized carbon particle (CP) solution or from CNP solution. A similar type of aggregation is observed by Iijima et al.[20] where small carbon particles are found to aggregate into ~ 80 nm size nano-horn structures. We have done some control SEM experiment to study this aggregation processes. Particle solution has been deposited on Si-substrate by a single or successive multiple drops (after the evaporation of first drop next drop was added). We have always found that multiple drop sample shows larger particle than the single drop. We have estimated the particle size of CNP as 12-15 nm for single drop sample (Figure 2c), but size increases when the sample was prepared from three drops (Figure 2d). This observation further indicates that the CNP agglomerate easily in solid form but remain isolated in water. AFM study of CNP also showed the existence of small particles as well as particle aggregates even if a very dilute solution was used for this study (supporting Figure S2). We have also estimated the particle size of CP and CNP from the dynamic light scattering (DLS) measurement and result shows that the hydrodynamic diameters is broad and ranges from 20 to 350 nm for CP but for CNP it shows a narrow distribution with average diameter of ~ 12.5 nm (supporting Figure S3). The broad size distribution and larger sizes in CP mask the presence of small carbon particle (CNP) in their size distribution. This broad DLS size distribution of CP and narrow size distribution of CNP corroborate the TEM observation. The increased average size of CNP in DLS study (in comparison to TEM which shows 2-6 nm) is because DLS consider overall hydrodynamic diameter that includes particle as well as adsorbed molecules and ions.

The soot contains mainly elemental carbon and oxygen having 96 atomic % and 4 atomic % respectively, whereas CNP shows that the C, O and N are 59 atomic %, 37 atomic % and 4 atomic % respectively as estimated from XPS compositional analysis. (Figure 3a-f), These data shows that CNP is mainly composed of graphitic carbon ($sp^2$) and oxygen/nitrogen bonded carbon, whereas starting soot is mainly composed of diamond like carbon ($sp^3$) with oxygen bonded carbon.[21-33] This composition variation is well matched with bonding structures obtained from FTIR spectrum (see supporting Figure S4 and identification of different bonding configurations).[34,35]

Figure 4 shows the Raman spectra of the CNP and soot respectively. Spectrum of CNP shows high photoluminescence background with compared to soot. The two signature peaks for carbon i.e. D band and G band are clearly seen for CNP and soot where D band corresponds to disordered structure in crystalline of $sp^2$ cluster and G band corresponds to the in-plane stretching vibration mode $E_{2g}$ of single crystal graphite. The intensity ratio ($I_D/I_G$), which is often used to correlate the structural purity to the graphite, also indicates that the CNP are composed of mainly nano-crystalline graphite.[35] The size of the nano-crystalline graphite obtained from the relation deduced by Ferrari et al.[35] was calculated as 2.2 nm. Similar graphitic structure is also obtained from x-ray diffraction patterns of CNP (supporting Figure S5).

It is well known that nitric acid oxidation produces OH and $CO_2H$ groups on the carbon nano-particle surfaces that made them hydrophilic and negatively charged particle.[36] In addition this oxidation can also induce small extent of nitration into graphitic carbon.[37] Our experimental data suggest that refluxing step with nitric acid has made two fold chemical modifications to the soot. First, it induces partial oxidation of carbons and introduces functional groups such as OH, $CO_2H$ and $NO_2$. Second, it induces doping of nitrogen and oxygen into the carbon particle. Introduction of functional groups induce water solubility and surface charge to the CNP. In addition it helps to break the large aggregated soot particle into small carbon particles. This oxidation step can also be considered as chemical route of incorporating nitrogen and oxygen into the carbon particle as observed from the chemical composition analysis.

The yield of soluble carbon particle depends of oxidation property of nitric acid in refluxing condition whereas the fluorescent quantum yield of carbon particle seems to depend on the efficiency of nitrogen and oxygen incorporation. Smaller particle size and dominant graphitic structure of the raw soot made this oxidation step easier. However, the efficiency of converting soot into soluble carbon particle is still low (yield ~20%), as observed from large part of insoluble soot. This suggests that this type of chemical oxidation is not efficient enough for complete conversion into water soluble particle. Longer time refluxing with nitric acid increases the yield of soluble particle to some extent but does not increase the yield of fluorescent carbon particle. This suggests that further oxidation might have other adverse effect such as further oxidation of carbon particle that reduces the conjugated double bond structure in the carbon particle.

Incorporation of nitrogen and oxygen defect via nitric acid oxidation might have a role in producing fluorescent centre into carbon particle. Such defect structures in the fluorescence property of diamond is well established.[1,3] As soot has some percentage of diamond like carbon (as observed from XPS data), it might happen that during oxidation step nitrogen and oxygen defects are formed into diamond structure. However, their presence in CNP is too low to determine with our presented analytical method and further study is needed to confirm this possibility. An alternative explanation of fluorescence may be that chemical oxidation and doping step introduces more conjugated double bond system into the carbon particle and thereby introduce the fluorescence property. Nevertheless, the advantage of this type of chemical processes of making fluorescent carbon particle is that it is simple and requires less adverse conditions as compared to ion beam radiation method.[1,3] However, as synthesized carbon particles have heterogeneous size distribution and small size particle are more fluorescent than larger one. Thus successful isolation of small particle is essential for the enhancement of fluorescence quantum yield.

Water soluble fluorescent carbon nanoparticle is ideal cell imaging probe with minimum cytotoxicity.[3,7,13] However, functionalization is an important step for cellular and sub-cellular targeting.[19] Interestingly, we found that small carbon particles (CNP) enter into cell without any further functionalization and using the fluorescence property of CNP it is possible track the CNP (Figure 5). CNP solution has been mixed with cell culture media along with cell, incubated for 30 minutes and washed cells were then imaged under bright filed, UV and blue excitations. Cells become bright blue-green under UV excitations and yellow under blue excitation, but they are colorless in the control sample where no CNP was used. This suggests that CNP enter into the cells and labeled cell can be imaged using conventional fluorescence microscope. MTT and Trypan blue assay of cell viability study suggest that CNP has no cytotoxicity. We exposed the cell with CNP of 0.1-1 mg/mL (which is about 100-1000 times higher than it required for imaging application) and for 24 hours. The cell survival rate in <0.5 mg/mL was between 90-100%, suggesting a minimum cell death. However, at higher concentration some percentage of cell death is observed. This result concludes that CNP can be used in high concentration for imaging or other biomedical applications.

## Conclusion

Fluorescent carbon nanoparticle of 2-6 nm in diameter was obtained after nitric acid oxidation of soot-particles. Surface oxidation and subsequent nitrogen and oxygen doping afforded light-emitting property of carbon particle. It is to be noted that the light emitted by these carbon particles depends on the wavelength of light used for excitation. We isolated different size particles and found that emission quantum yield is size-dependent, i.e. smaller the size better is their photoluminescence efficiency. Our approach can be used for milligram scale synthesis of these water soluble particles. The fluorescence property of these particles is useful for cell imaging application. These CNP enter into cell without any further functionalization and fluorescence property of the particle can be used to track their position in cell using conventional fluorescence microscope. The discovery of fluorescent carbon nanoparticles will no doubt lead to more research in this filed, particularly these particles have the potential in biomedical application where cadmium-based quantum dot shows toxic effects. However, synthetic methods of these particles need to be much more advanced so that large quantity of these particles with different emission color can be easily prepared.

**Acknowledgement**

The authors would like to thank Dr. Nihar **R.** Jana of National Brain Research Centre (NBRC), Gurgaon, India for providing cellular imaging facility and cytotoxicity study. Authors would like to thank DST (SR/S5/NM-47/2005), Government of India for providing financial support. AS thanks to CSIR, India for providing fellowship.

**Supporting Information Available:** Fluorescence spectra of CP produced at different reflux time, FTIR spectrum and AFM image of CNP. These materials are available free of charge via the Internet at http://pubs.acs.org

# Scheme and Figures

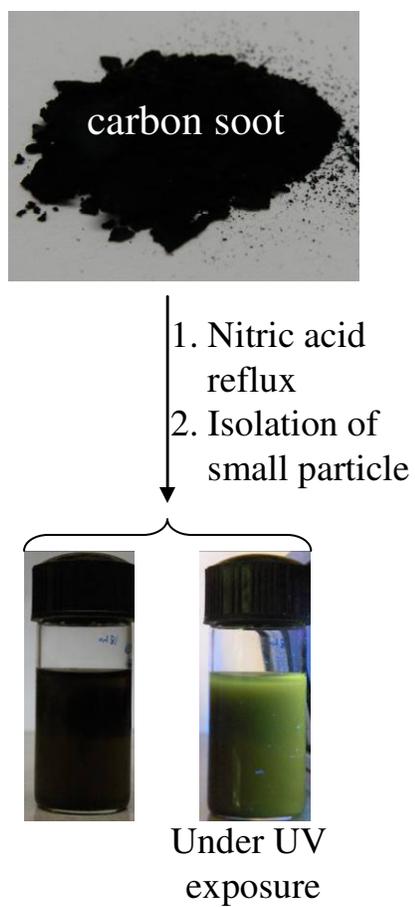

**Scheme 1:** Steps in the preparation of fluorescent carbon nanoparticle (CNP) from soot.

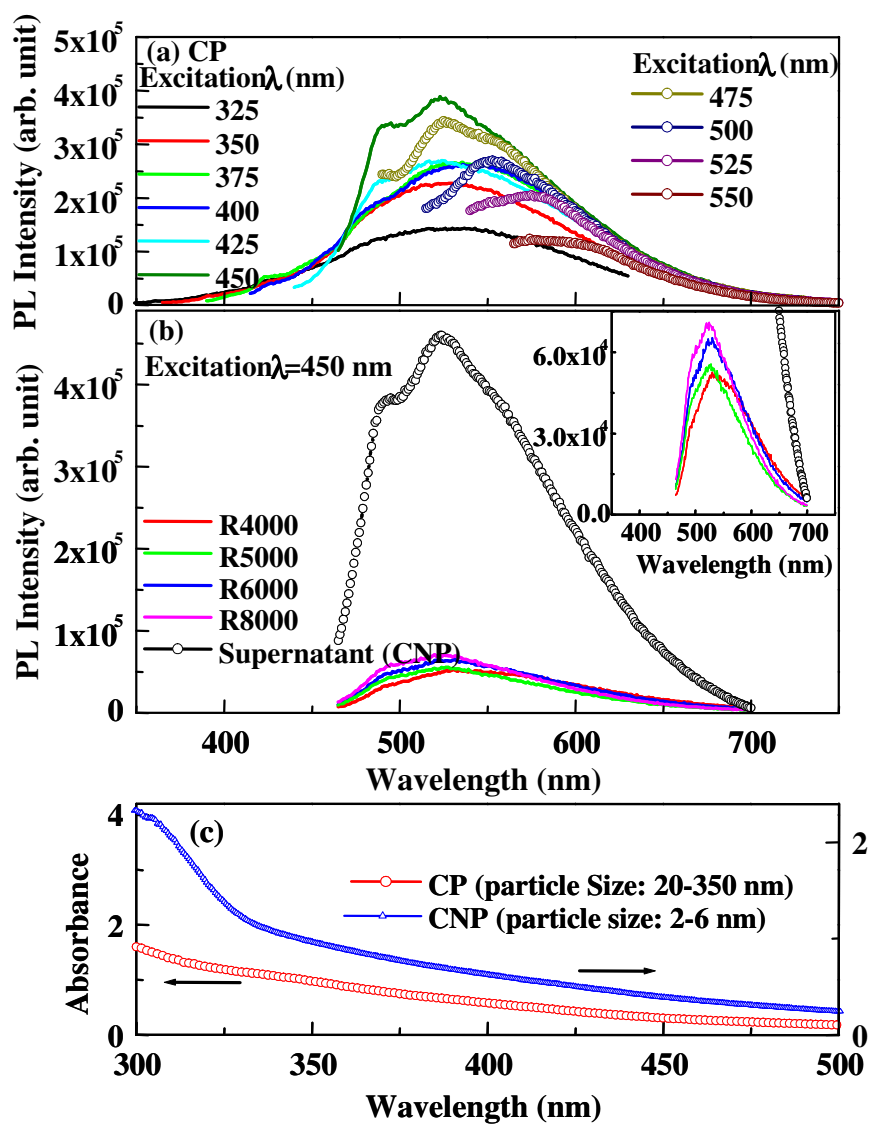

Figure 1. (a) Fluorescence spectra of as synthesized aqueous carbon particle (CP), obtained by 12 hrs refluxing, excited at different wavelength. (b) Fluorescence spectra of different size CP showing the most intense spectrum for smallest size (shown as supernatant which corresponds to CNP of 2-6 nm). As synthesized CP solution was dissolved in 2 mL of water-ethanol-chloroform solution and then centrifuged at different speed. After each step of centrifugation the precipitate was dissolved in 2 mL of fresh water. The smallest size CP (shown as supernatant) does not precipitate even at 16000 rpm. All the solutions are excited at 450 nm. (c) UV-visible absorption spectra of aqueous solution of as synthesized carbon particle (CP) and smallest size carbon particle of 2-6 nm (CNP).

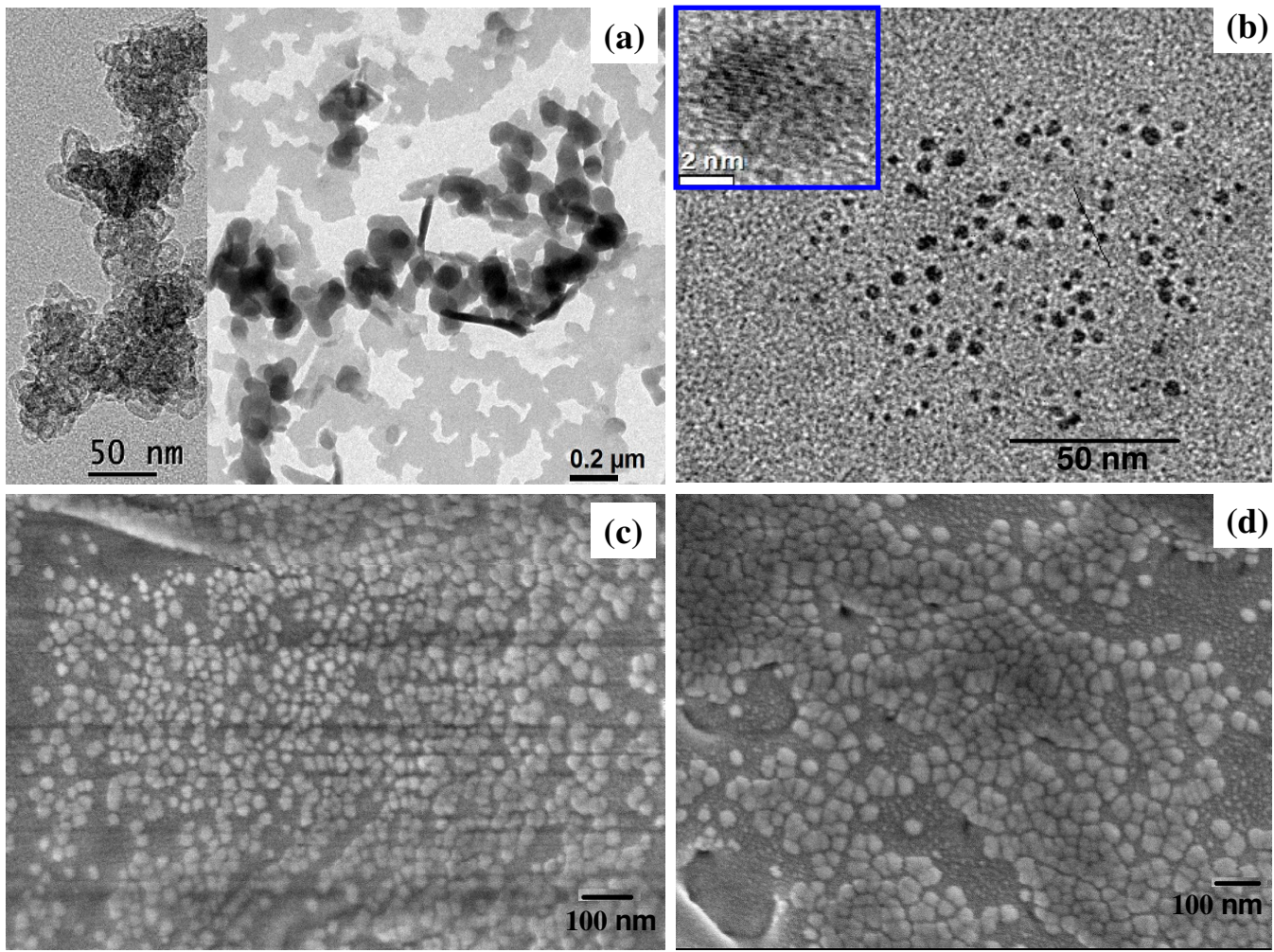

Figure 2. (a) Transmission electron microscopy (TEM) images of as synthesized carbon particle (CP) showing broad size/shape distribution as well as extensive particle agglomeration. (b) TEM images of small size carbon particle (CNP) with high resolution image of one particle at the inset. (c) Scanning electron microscopy image of single drop of CNP solution onto Si-substrate; (d) Scanning electron microscopy image of CNP solution on Si-substrate but after having three successive drops. It shows that particle sizes increases due to agglomeration.

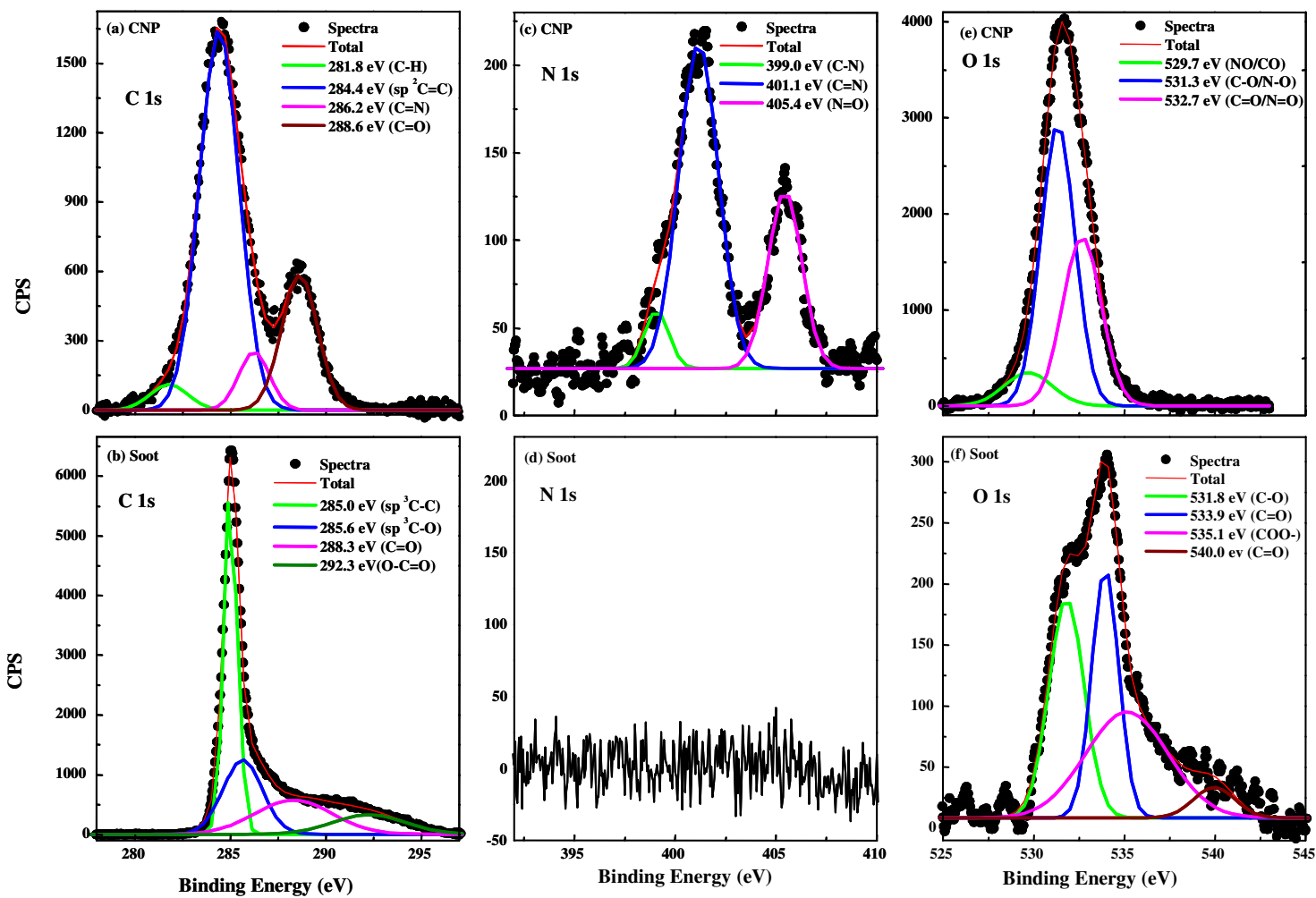

Figure 3. X-ray photoelectron spectroscopy (XPS) spectra of (a) C 1s (c) N 1s and (e) O 1s of CNP solution deposited on glass substrate and (b) C 1s (d) N 1s and (f) O 1s of raw candle soot (powder). XPS compositional analysis shows a significant change in the percentage distribution of elements; from 96% C and 4% O in raw candle soot into 59% C, 37% O and 4% N in the CNP.

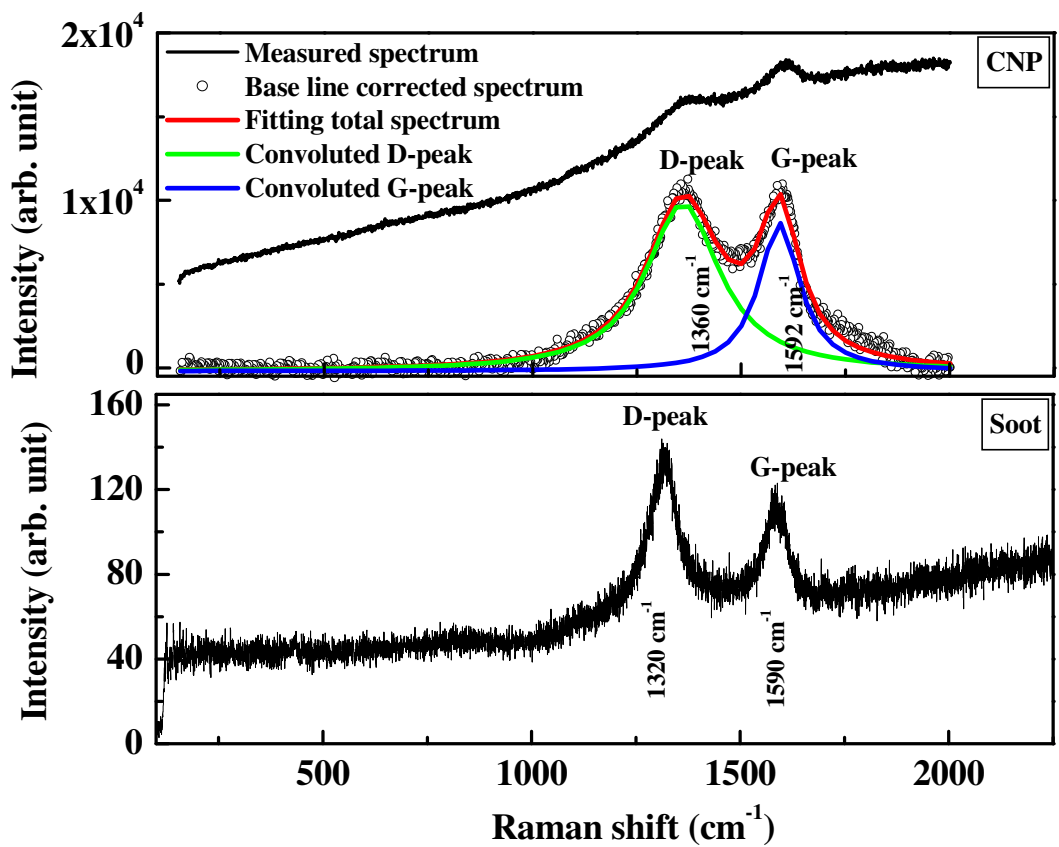

Figure 4. Raman spectra of CNP and raw soot (powder). Fitting curve of base line corrected spectrum of CNP shows the clear D and G peaks and their intensity ratio i.e. ($I_D/I_G$) is equal to 2 indicates the structure of CNP is nano-crystalline graphite.

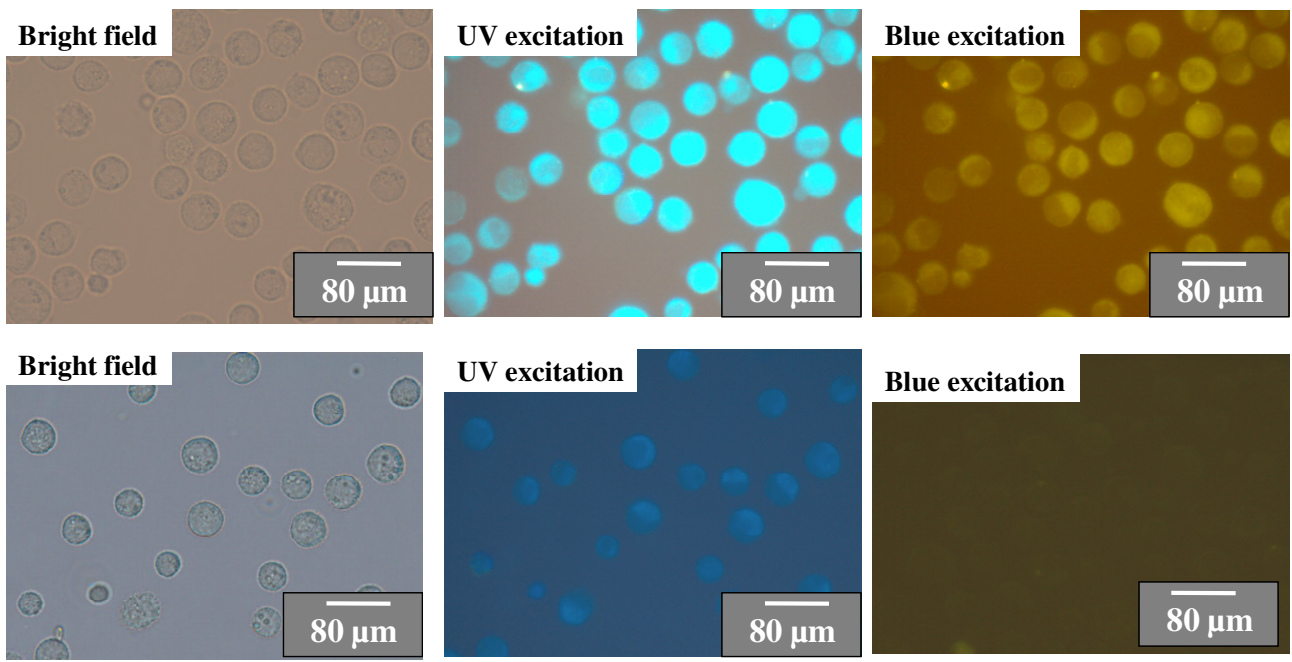

Figure 5. Fluorescent CNP based labeling of EAC cells. Cells solution was mixed with CNP solution and incubated for 30 minutes. Washed cells were imaged under bright filed, UV and blue excitations. The bottom row images correspond to the control experiment where no CNP was used. Cells become bright blue-green under UV excitations and yellow under blue excitation, but they were colorless in the control sample. A light blue color of the control sample under UV excitation is due to well known auto-fluorescence of cell.

**Table of Contents (TOC)**

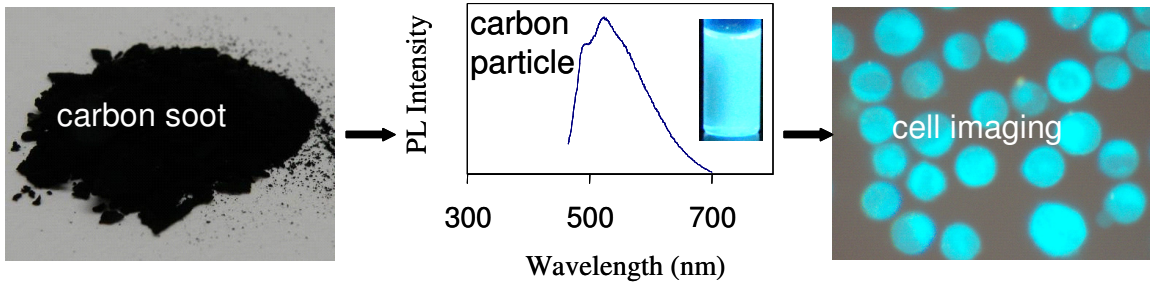

# Supporting Information

# Fluorescent Carbon Nanoparticle:
# Synthesis, Characterization and Bio-imaging Application

S.C. Ray[a],*, Arindam Saha, Nikhil **R.** Jana* and Rupa Sarka**r**

*Centre for Advanced Materials, Indian Association for the Cultivation of Science, Kolkata-700032 (India)*

**XPS peak assignment:** XPS C-1*s*, N-1*s* and O-1s spectra and deconvolution of these spectra into different peaks was executed by curve fitting using Gaussian functions. C-1*s* peak of CNP is consisted of four Gaussian peaks (Figure 4a) centered at 281.8 eV, 284.4 eV, 286.2 eV and 288.6 eV. The peak at 284.4 eV indicates that carbon is mostly in the form of graphite and assigned to *sp*$^2$ aromatic hydrocarbons.[21,22] The peak at 286.2 eV can be assigned to C atoms surrounded by N and H atoms i.e. C-N / C=N and C-H bonds[23-25] and the peak at 288.6 eV is the -C=O carbonyl groups.[26,27] The peak at lower energy 281.8 eV is presumably the C-H peak. In case of soot the peaks in C-1s are consisted of *sp*$^3$ C-C (285.0 eV)[28], C-O/C-H (285.6 eV), carbonyl groups (288.4 eV)[28] and the peak at 292.3 eV can be attributed to $CO_2$ and/or C-C=O bonds.[29] In case of N 1s (Figure 4b), the peak centered at 399 eV and 401.1 eV are graphitic structured C-N and C=N bond respectively[30-32], whereas the peak at 405.4 eV is assigned to some oxidized N-species like N-O and/or N=O bonds.[28] Meanwhile, in case of CNP, the O1*s* peak consisted of three Gaussian peaks (Figure 4e) centered at 529.7 eV, 531.3 eV and 532.7 eV, which are associated with C-O/N-O and C=O/N=O respectively. In case of soot, the O1s peak is consisted of four Gaussian peaks and are associated with C-O (531.8 eV), -C=O (533.9 eV) and COOH (535.1 eV and 540.0 eV) respectively.[33]

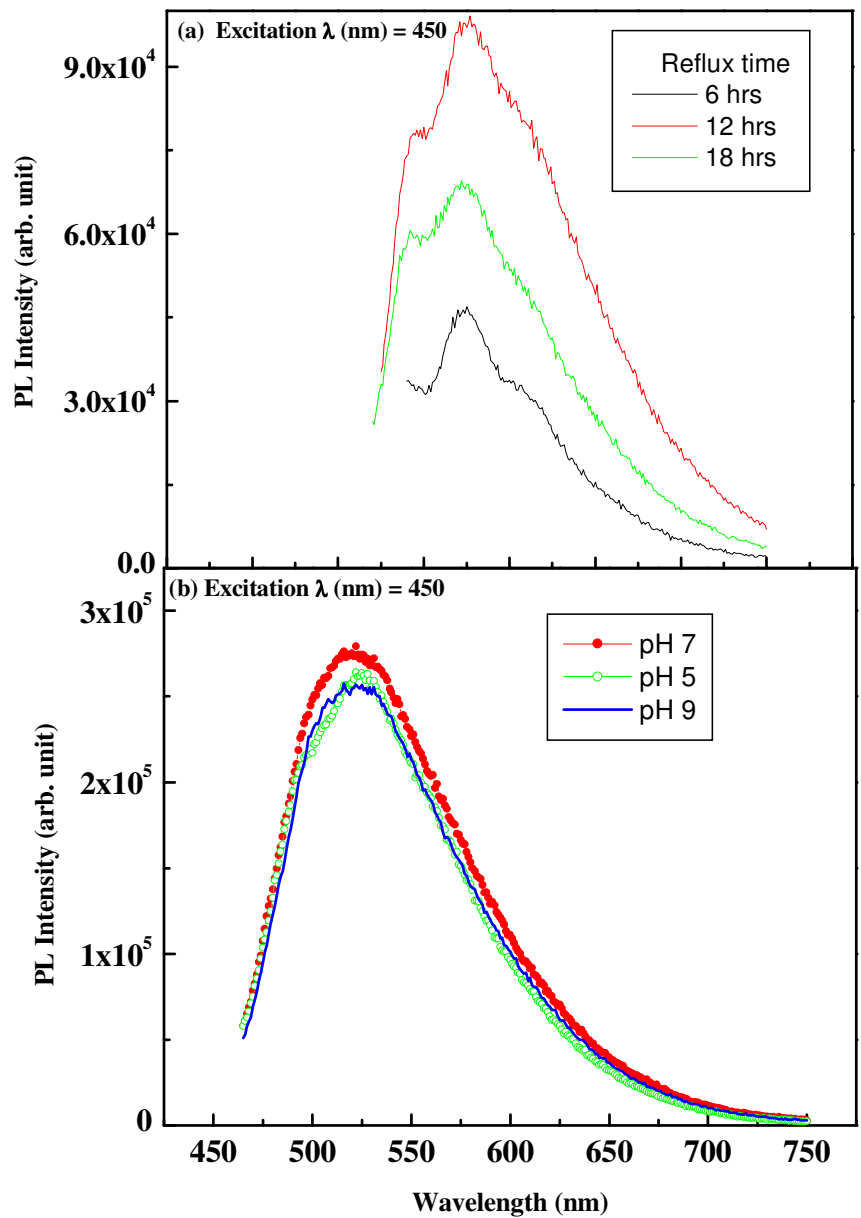

Figure S1. (a) Fluorescence spectra of carbon particle (CP) produced with different reflux time. (b) Fluorescence spectra of small size carbon particle of 2-6 nm (CNP) at phosphate buffer of different pH.

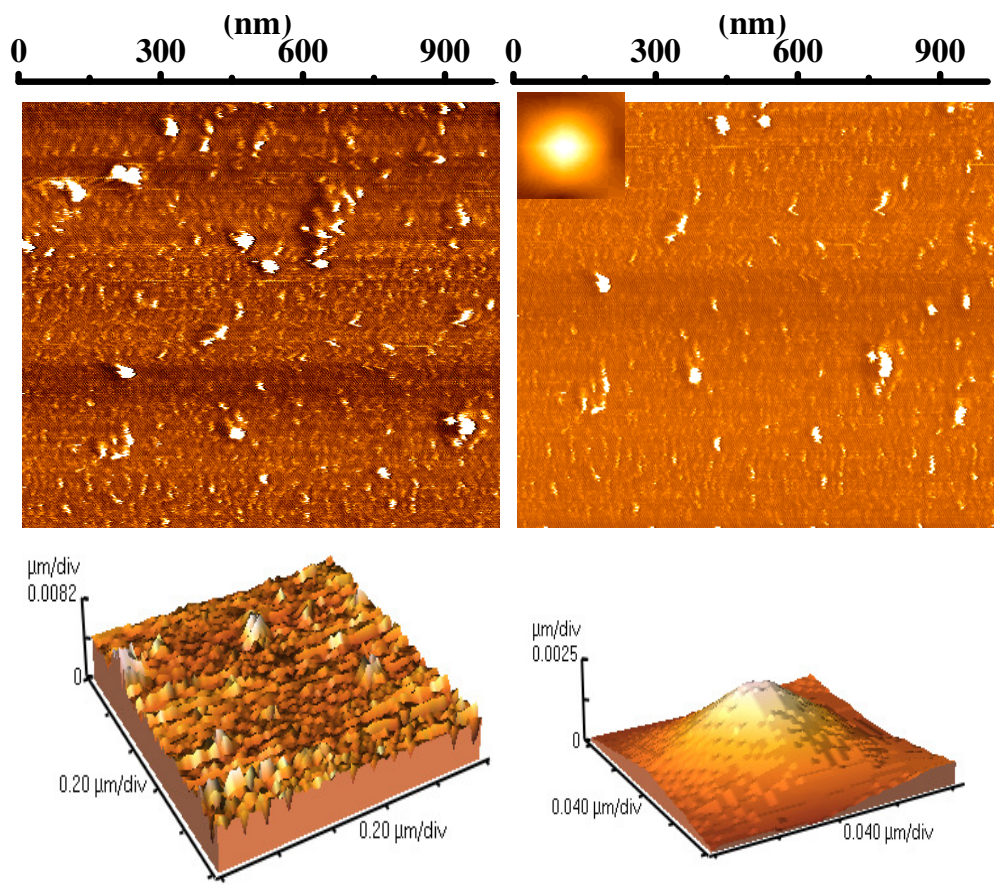

Figure S2. AFM image of CNP taken at different position and magnification.

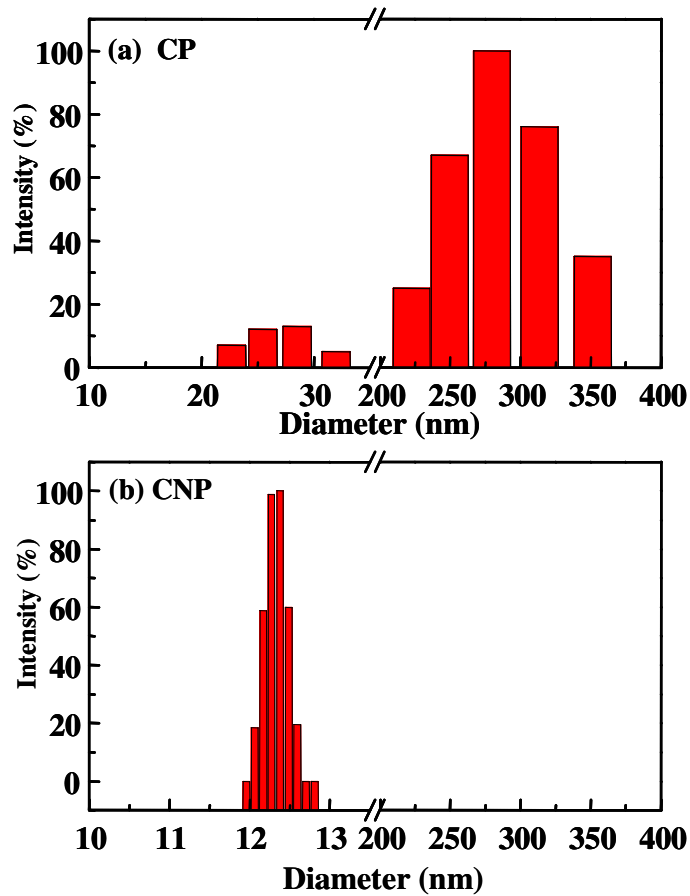

Figure S3. Particle size distribution of (a) aqueous solution of as synthesized carbon particle (CP) and (b) smallest size carbon particle of 2-6 nm (CNP) as observed from dynamic light scattering study.

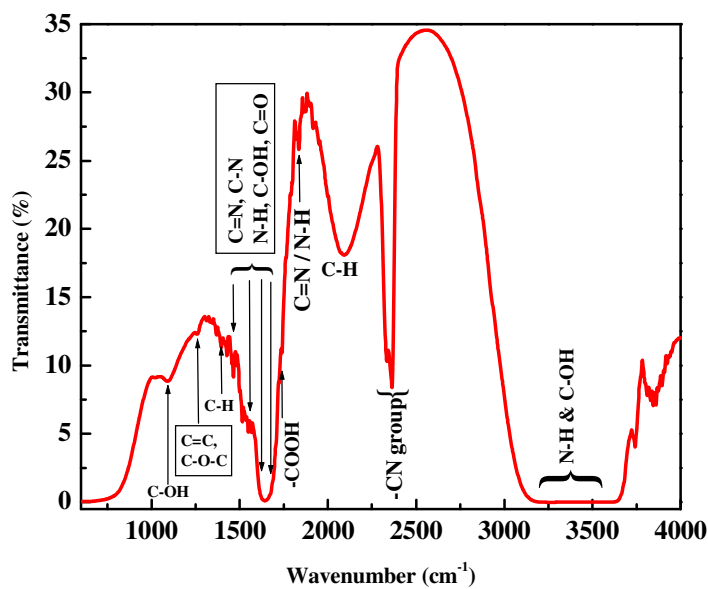

Figure S4. FTIR spectrum of CNP showing the presence of different functional groups. The peaks at ~1100 cm$^{-1}$ and 1265 cm$^{-1}$ are ascribed as C-O-C and C=C bonds respectively. A few small peaks are observed in between 1400–1650 cm$^{-1}$ and at 1735 cm$^{-1}$ ascribed to C-H, conjugated C-N and C=N, N-O bonds of stretching modes respectively. The broad band observed at 1600–1700 cm$^{-1}$ can be assigned to the C-OH and C=O bonds. A small band at 1835 is observed and is ascribed as associated with C=N (sp$^2$ C–N) stretching vibration. The band observed around 2090 cm$^{-1}$ is ascribed to C-H bond. The band 2335 and 2360 cm$^{-1}$ can be attributed to the nitrile (–CN) group. In addition a broad IR band ~ 3100–3650 cm$^{-1}$ appeared due to C-OH and COOH bonds.

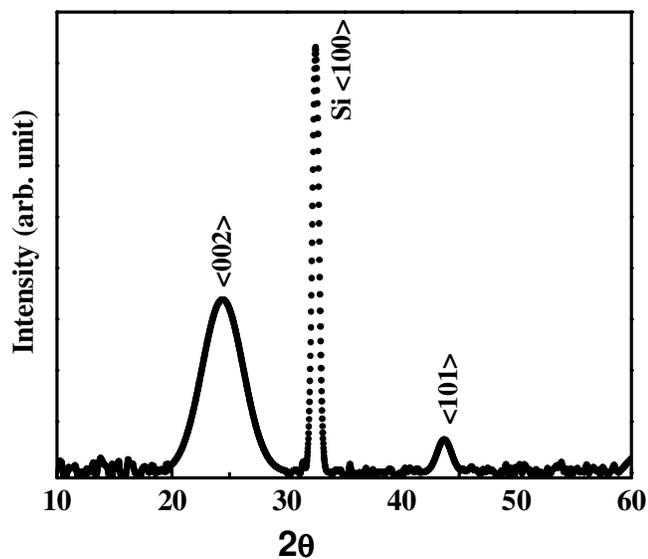

Figure S5. The wide-angle region of the x-ray diffraction (XRD, Shimadzu, XRD-6000) patterns of CNP exhibit a high-intensity diffraction peak at $2\theta = 24.4^0$ and one additional peat at $2\theta = 43.7^0$ that are ascribed to (002) and (101) diffraction of graphitic carbon respectively. Si peak is arises because CNPs are deposited on Si substrate using solution drop process and dried at room temperature for this measurement.

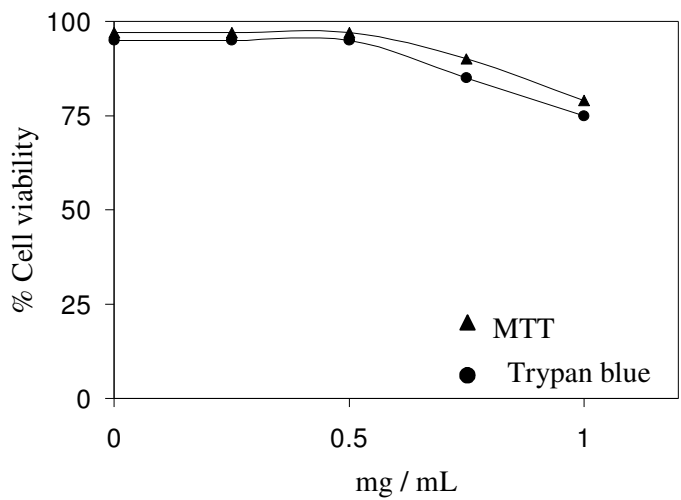

Figure S6. Cytotoxicity of CNP for HepG2 cell studied by MTT and Trypan blue assays. All data are average of three experiments.